\documentclass[fleqn,12pt,twoside]{article}
\usepackage{espcrc1}
\usepackage{graphicx}
\usepackage{bm}
\usepackage{dcolumn}

\usepackage{graphicx}
\usepackage[figuresright]{rotating}

\newcommand{\AmS}{{\protect\the\textfont2
  A\kern-.1667em\lower.5ex\hbox{M}\kern-.125emS}}

\title{Strange vector form factors in the context of 
the SAMPLE, A4, HAPPEX and G0 experiments}
\author{Antonio Silva\address[a]{
Institut f\"ur Theoretische  Physik  II,   
Ruhr-Universit\" at Bochum,\\ 
 D--44780 Bochum, Germany}\address{
Departamento de F\'\i sica and Centro de F\'\i sica Computacional,\\
Universidade de Coimbra,
P-3000 Coimbra, Portugal},
\underline{Hyun-Chul Kim}\address{Department of Physics,
Pusan National University,\\
609-735 Pusan, Republic of Korea}
and Klaus Goeke\addressmark[a]}

\begin{document}
\maketitle

\begin{abstract}
We present the recent results of the 
strange vector form factors  of the nucleon within the framework 
of the SU(3) chiral quark-soliton model.  We compare our results 
with the recent experimental data of the SAMPLE
and HAPPEX collaborations and find that they are in a good 
agreement with the data.  We also predict the future experiments of the
A4, HAPPEX-II and G0 collaborations.
\end{abstract}
\vspace{1.cm}

The strange vector form factors of the nucleon have been a hot
issue recently, as their first measurement was conducted by the SAMPLE
collaboration~\cite{Mueller:1997mt,Spayde:2000qg,SAMPLE00s} at
MIT/Bates, parity-violating electron scattering being used.  
The HAPPEX collaboration at JLAB also announced the measurement of 
the strange vector form factors~\cite{Aniol:2000at}.  The A4
collaboration at MAMI in Mainz will soon bring out new measurements
and the HAPPEX-II and G0 experiments at JLAB are being preprared. 

There has been a great deal of theoretical attempts to predict the
strange vector form factors, differing by the methods employed to 
investigate the non-perturbative physics behind them.  In a recent
review~\cite{Beck:2001dz}, many of theoretical works were reviewed.

We shall use the chiral quark-soliton model
($\chi$QSM) as our theoretical framework.   
It is an effective quark theory based on the instanton vacuum of QCD 
and results in a Lagrangian for valence and sea quarks interacting 
via a static self-consistent pseudo-Goldstone background field.  
The $\chi$QSM has been successful in describing the electromagnetic and 
axial-vector form factors~\cite{Christov:1996vm} in SU(2) 
and the forward and generalized parton distributions
\cite{Diakonov:1996sr,Petrov:1998kf,Goeke:2001tz}.   

Recently, we have studied the strange form factors
in the context of the SAMPLE, HAPPEX, HAPPEX-II, A4 and G0 experiments 
within the framework of this $\chi$QSM~\cite{Silva:2001st,Silva:2002ej},
employing symmetry-conserving quantization~\cite{Praszalowicz:1998jm} 
and taking into account
kaonic effects.  In this talk, we want to present the recent
results on the strange form factors and discuss their aspects 
in comparison with those experiments.  

The results for the strange electric form factor $G_{E}^{\rm s}$ are 
shown in Fig.~1 and for the strange magnetic form factor 
$G_{M}^{\rm s}$ in the left panel of Fig.~2.  The neutral vector form
factor is drawn in the right panel of Fig.~2.  
\begin{figure}\centerline{\includegraphics[height=5cm]{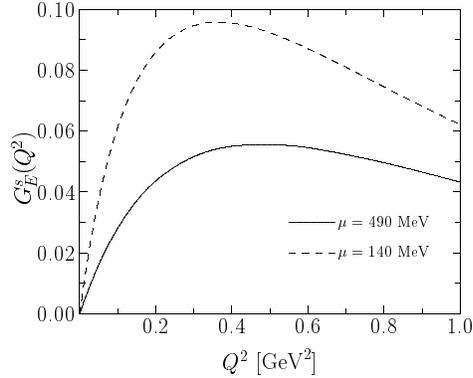}}
\caption{The strange electric form factor $G_E^{\rm s}$ as a function 
of $Q^2$ with the kaon ($\mu=490$ MeV) and pion ($\mu=140$ MeV) 
asymptotic tails.  The constituent quark mass is $M=420$ MeV.}
\end{figure}
\begin{figure}
  \begin{minipage}[h]{6.2cm}
\centerline{\includegraphics[height=5.0cm]{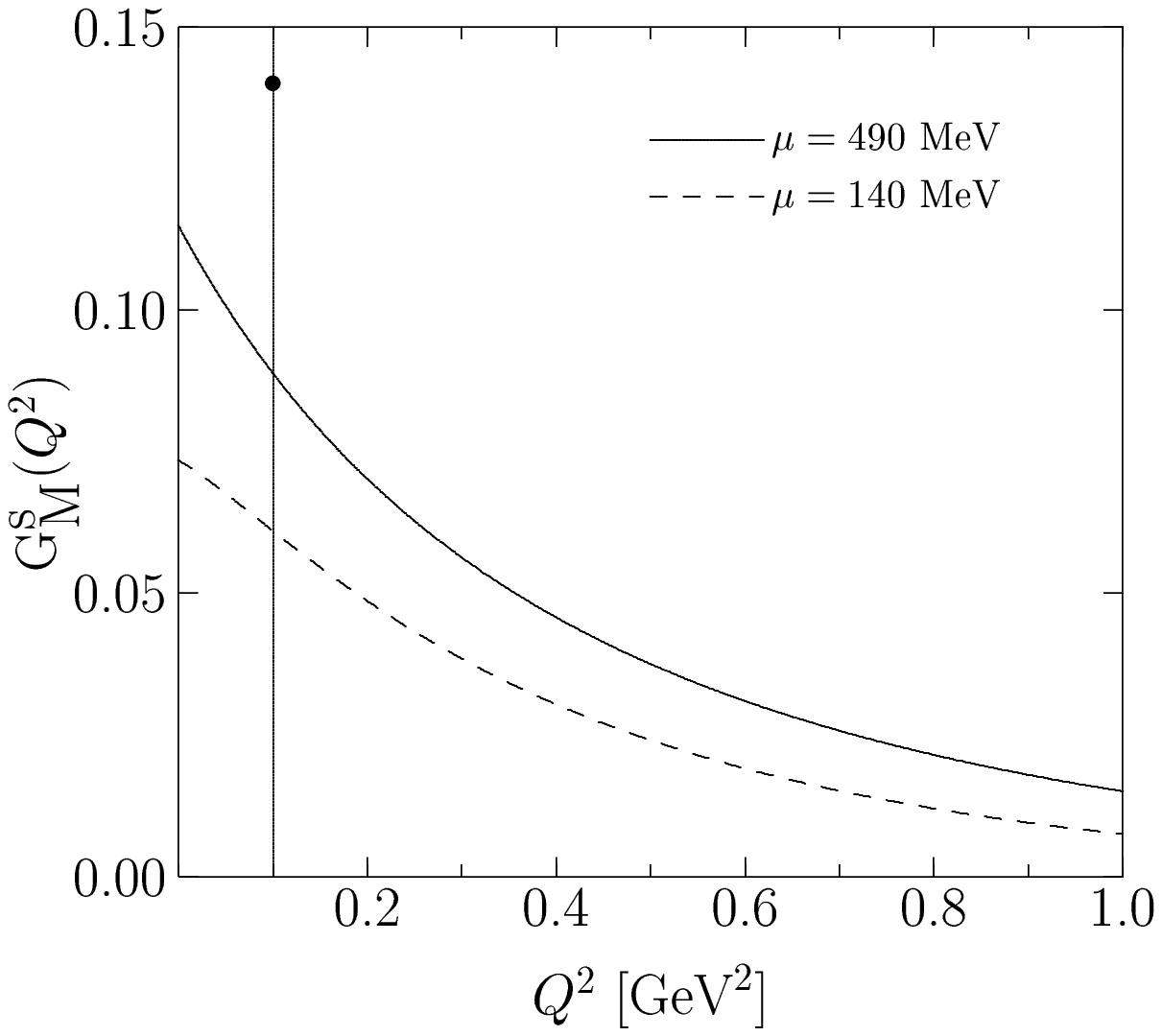}}
  \end{minipage}\hspace{2cm}
  \begin{minipage}[h]{6.2cm}
\centerline{\includegraphics[height=5.0cm]{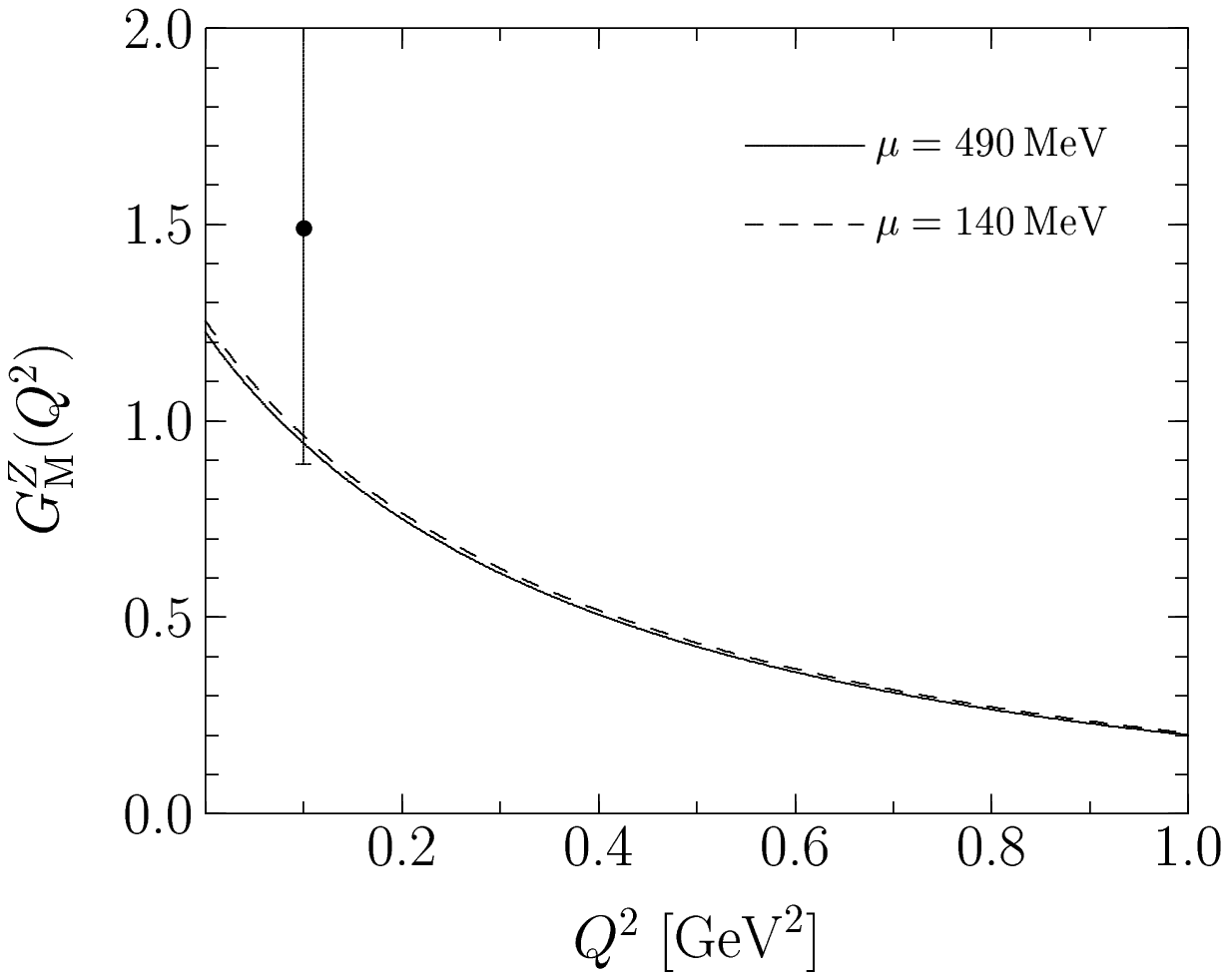}}
  \end{minipage}
\caption{The strange magnetic form factor $G_M^{\rm s}$ and the
  neutral vector form factor in units of $\mu_N$
         as a function of $Q^2$.  
         Conventions and model parameter as in Fig.~1.  The
         experimental data are taken from the SAMPLE  
experiment~\cite{SAMPLE00s}.}
\end{figure}
As shown in Fig.~2, the model results are very close to the SAMPLE 
data, but the experimental error is too large to draw any conclusion as yet.
These figures show a general effect due to the influence of the 
Yukawa mass $\mu$, governing the asymptotic behavior of the profile 
function at large $r$, $\exp(-\mu r)/r$, on the form factors.  
It is expected that the kaon influence on the asymtotics of the 
meson fields play an important role in the description of the 
strange vector form factors, since strange quarks in the nucleon 
may be excited in the form of virtual kaons and hyperons. 
Keeping this in mind, we implement a heavier mass asymptotic 
behavior of the profile function of the pseudo-Goldstone soliton
field.  We shall regard the difference arising from two different
profile functions as our systematic uncertainty of the model.

The results for the strange electric and magnetic radii are:
$\langle r^2\rangle_{E}^{\rm s}=-0.220 (\pi)\sim -0.095 (K)$ fm and 
$\langle r^2\rangle_{M}^{\rm s}=0.303 (\pi)\sim 0.631 (K)$ fm,
respectively.  Those for the strange magnetic moment are: 
$\mu_{\rm s}=0.074 (\pi)\sim 0.115 (K)\,\mu_N$. 

\begin{table}[t]\begin{center}\begin{tabular}{|c|c|} \hline              
$(G^0_E+0.392 G^0_M)/(G^p_M/\mu_p)$ & $G^{\rm s}_E+0.392G^{\rm s}_M$  
\\ \hline
$1.433\sim 1.695$ $(\mu = m_{\pi}\sim m_{\rm K})$ &  
$0.103 \sim 0.071$ $(\mu = m_{\pi}\sim m_{\rm K})$  \\ \hline
$1.527\pm0.048\pm0.027\pm0.011$(HAPPEX)  &
$0.025\pm0.020\pm0.014$ (HAPPEX)\\ \hline     
\end{tabular}\end{center}
\caption{The combinations $(G^0_E+0.392G^0_M)/(G^{p\gamma}_M/\mu_p)$ and 
$G^{\rm s}_E+0.392G_M^{\rm s}$ for the HAPPEX kinematics $Q^2=0.477$ 
${\rm GeV}^2$.   The model parameters as in Fig.~1.
The experimental data are taken from HAPPEX~\cite{Aniol:2000at}.}
\end{table}

In Table 1 we present the results relevant for the HAPPEX
experiments.  While the model result of the flavor-singlet form factor
is in a fairly good agreement with the HAPPEX data, that of the
strange vector one shows a little bit higher value than the
experimental one.  In Tables 2-4, we list the predictions of the 
future experiments.
 
\begin{table}[ht]
\caption{Strange form factors: The prediction for the G0 experiment.
The model parameters as in Fig.~1.}
\begin{tabular}{|c|c|c|c|c|}
\hline
 & \multicolumn{2}{c|}{$\theta=10^\circ$ }
& \multicolumn{2}{c|}{$\theta=108^\circ$}
\\ \cline{1-5}
$Q^2\;[{\rm GeV}^2]$ & $\beta$ & $G_E^{{\rm s}} + \beta G_M^{{\rm
    s}}$ $(\mu = m_{\pi}\sim m_{\rm K})$ & $\beta$ 
&  $G_E^{{\rm s}} + \beta G_M^{{\rm s}}$ 
$(\mu = m_{\pi}\sim m_{\rm K})$
\\ \hline
$0.16$  & $0.13$ & $0.09\sim 0.05$ & $0.63$ & $0.11\sim 0.09$ \\
$0.24$  & $0.20$ & $0.10\sim 0.06$ & $0.99$ & $0.14\sim 0.11$ \\
$0.325$ & $0.26$ & $0.11\sim 0.07$ & $1.31$ & $0.14\sim 0.13$ \\
$0.435$ & $0.35$ & $0.11\sim 0.07$ & $1.81$ & $0.15\sim 0.14$ \\
$0.576$ & $0.47$ & $0.10\sim 0.07$ & $2.49$ & $0.14\sim 0.14$ \\
$0.751$ & $0.61$ & $0.08\sim 0.06$ & $3.35$ & $0.12\sim 0.13$ \\
$0.951$ & $0.81$ & $0.07\sim 0.06$ & $4.62$ & $0.11\sim 0.12$ \\
\hline
\end{tabular}
\end{table}

\begin{table}[ht]
\caption{Strange form factors: The prediction for the A4 experiment.  
The model parameters as in Fig.~1.}
\begin{tabular}{|c|c|c|c|c|}
\hline
 & \multicolumn{2}{c|}{$\theta=35^\circ$ }
& \multicolumn{2}{c|}{$\theta=145^\circ$}
\\ \cline{1-5}
$Q^2\;[{\rm GeV}^2]$ & $\beta$ & $G_E^{{\rm s}} + \beta G_M^{{\rm
    s}}$ $(\mu = m_{\pi}\sim m_{\rm K})$& $\beta$ 
&  $G_E^{{\rm s}} + \beta G_M^{{\rm s}}$
$(\mu = m_{\pi}\sim m_{\rm K})$
\\ \hline
$0.10$  & $0.099$ & $0.07 \sim 0.04$ & $-$ & $-$ \\
$0.227$ & $0.22$ & $0.10\sim 0.06$ & $4.07$ & $0.28\sim 0.32$ \\
$0.47$ & $-$ & $-$ & $8.963$ & $0.33\sim 0.42$ \\
\hline
\end{tabular}
\end{table}

\begin{table}[ht]
\caption{Strange form factors: The prediction for the HAPPEX II experiment.  
The model parameters as in Fig.~1.}
\begin{tabular}{|c|c|c|}
\hline
 & \multicolumn{2}{c|}{$\theta=6^\circ$} 
\\ \cline{1-3}
$Q^2\;[{\rm GeV}^2]$ & $\beta$ & $G_E^{{\rm s}} + 
\beta G_M^{{\rm s}}$ $(\mu = m_{\pi}\sim m_{\rm K})$
\\ \hline
$0.11$  & $0.09$ & $0.07\sim 0.04$ \\
\hline
\end{tabular}
\end{table}
In this talk, our aim has been to present the recent results of 
the strange form factors of the nucleon within the framework of
the SU(3) chiral quark-soliton model.  We have taken into account 
the new scheme of the SU(3) quantization.  
We also have considered two different Yukawa asymptotic tails, that
is, the pion and kaon tails.  The difference between the results with
the pion and kaon tails is regarded as the uncertainty which is 
inherent in the present model.  The results are in a fairly good 
agreement with the SAMPLE and HAPPEX results.  We also predict the
future experiments of the A4, HAPPEX-II and G0 experiments.

AS acknowledges partial financial support from Praxis XXI/BD/15681/98.
The work of HCK is supported by the KOSEF (R01--2001--00014).  The
work has also been supported by the BMBF, the DFG, 
the COSY--Project (J\" ulich) and POCTI (MCT-Portugal).

\end{document}